# Electrical glassy behavior in granular aluminium thin films

Julien Delahaye, Thierry Grenet

*Institut Néel, CNRS-UJF, BP166, 38042 Grenoble, France*

**Abstract**

We present new results obtained by field effect measurements on insulating granular Al thin films. First, reproducible and stable conductance fluctuations are seen in micron size samples as a function of gate voltage. The anomalous field effect and its slow relaxation already known to exist in macroscopic samples are shown to still exist in small samples and to have no influence on the fluctuations pattern. Secondly, "true" aging is demonstrated, i.e. the anomalous field effect relaxation depends on the time elapsed since the cooling, the longer this time the longer it takes for the system to react to a gate voltage change. Interpretations and implications of these findings are discussed.

## 1. Introduction

The interest in the electronic properties of disordered insulators has been revived by the field effects measurements performed by Ovadyahu et al. on indium oxide thin films [1]. At liquid helium temperature, the conductance of the films was found to increase when the gate voltage $V_g$ was swept away from its cooling value, whatever the sweep direction ("anomalous" field effect or conductance dip). Moreover, at constant $V_g$, the conductance was shown to decrease as a logarithm of time after a temperature quench or a sudden gate voltage change. This decrease which corresponds to the formation of the conductance dip was observed over more than 5 orders of magnitude in time. Qualitatively similar effects were reported in other systems, like discontinuous and ultra-thin films [2], and we published recently a thorough study of these phenomena in granular aluminium films [3]. There were also rare reports of slow conductance relaxations in doped semiconductors [4] (limited to temperatures below 1K) and in 2D electron systems [5].

The fact that these slow relaxation phenomena are mostly observed at 4K in disordered insulators having a high charge carrier density raises the question whether they reflect the existence of the Coulomb glass theoretically predicted more than 20 years ago [6]. Such an interpretation is strongly supported by doping effects in indium oxide films [7], where it was shown that the slow conductance relaxation exists only for carrier densities higher than $10^{19}cm^{-3}$. In an alternative explanation developed initially for granular systems [8], such effects could stem from a slow response of atomic or ionic configurations close to the conducting paths [3,9]. In spite of an increasing number of experimental and theoretical studies on this topic, the interpretation of the anomalous field effect and its slow relaxation remains an open question.

We aim here to discuss two aspects of the glassy effects in disordered insulators, in the light of recent experiments on granular Al thin films. In section 3, we present how the conductance dip coexists with mesoscopic conductance fluctuations in small size samples. This is especially interesting since the conductance dip in macroscopic samples was found to be associated with the slow conductance relaxations, whereas the mesoscopic conductance fluctuations visible in small size samples are the fingerprint of the microscopic disorder landscape. In section 4, we report new experimental procedures that reveal for the first time a “true” ageing in the gate voltage induced conductance relaxations.

## 2. Experimental

We measured the conductance of MOSFET-like devices in which the conducting channels are made of insulating granular Al thin films. Granular Al films 20nm thick are deposited by e-beam evaporation of Al under a controlled pressure of oxygen. The electrical resistance per square $R_{\square}$ of the films at 4K can be tuned between few MΩ to few GΩ by slightly changing the oxygen pressure. The films are believed to consist of Al grains with typical diameters of 5nm, separated by thin insulating $AlO_x$ layers allowing tunneling between them. Gate and gate insulator (100nm thick) materials are either $Si_{++}$ / $SiO_2$ or Al / $AlO_x$.

## 3. Mesoscopic conductance fluctuations

In small size samples, reproducible conductance fluctuations are visible as a function of the gate voltage $V_g$. Typical relative amplitude is 0.1% for a sample of size 30μm x 30μm. The conductance fluctuations pattern remains almost unchanged during many days after a quench at 4K. Details of the pattern are sample-dependent and specific to each cool down. No periodicity or characteristic gate voltage scale has been identified.

The conductance values distribution is Gaussian like and its relative root mean square amplitude decreases like the inverse of the square root of the films area. This indicates that the fluctuations result from the sum of larger and statistically independent microscopic fluctuations. A detailed study of the $R_{\square}$ dependence reveals a good agreement with the percolation theory predictions applied to strongly disordered media [10]. In the presence of an exponentially large distribution of microscopic resistances (which is quite common for hopping disordered insulators), the conductance starts to self average only above a characteristic length scale called the homogeneity length that can be significantly larger than the coherence length. In our case, clear fluctuations are still visible even in $mm^2$ size samples. These fluctuations, like the Universal Conductance Fluctuations in diffusive metals, reflect the sample specific realization of disorder. In our granular Al films, the exponentially large distribution of microscopic resistances comes from the combined effects of a charging energy $E_C$ larger than the thermal energy ($E_C \approx$ 150K) and an important potential disorder on the grains (distribution of island sizes and shapes, broad distribution of offset charges…).

In not too small samples, we can resolve both the conductance fluctuations and the conductance dip. This allows us to determine if the formation (or the erasing) of a conductance dip is accompanied by significant changes in the fluctuations pattern. Our conclusion is that, within our measurement accuracy, the two phenomena appear to be independent of each other, like two additive contributions to the conductance [10]. The conductance fluctuations remain unchanged under the formation of a conductance dip (see Fig. 1). And conversely, the conductance dip parameters (width, amplitude and relaxation functions) are the same whatever the conductance fluctuations amplitude is, at least down to sizes of 20μm.

We can draw two conclusions from this independence. First, the conductance dip does not result from an incomplete averaging of some random microscopic conductance modulation. Second, its formation does not induce major changes in the disorder potential landscape. A simple interpretation would be that the conductance dip is of electronic origin whereas the disordered potential (printed in the conductance fluctuations pattern) is fixed by ionic and atomic configurations. But as we discuss in details in [10], the conductance fluctuations are also certainly influenced by the electronic interactions between the Al grains. The independence between the two phenomena could then simply result from the fact that the conductance dip corresponds to small changes of the macroscopic conductance (no more than few percent in our case) and is not able to affect significantly the microscopic hopping parameters. Similar results were also found recently on indium oxide thin films [11].

## 4. "True" aging

Beyond the question of the conductance dip origin, it is important to investigate the glassy properties of our granular Al films and to compare them with other glasses. Glasses include quite different systems, like structural glasses, spin glasses, gels, colloids, etc. They share characteristic properties such as a large distribution of relaxation times (responsible for non exponential slow relaxations) and the so called "aging" effects. Aging refers to the singular property that the response of a glassy system depends on its "age", i.e. on the time elapsed since cooling. It originates from the growth of correlations as the system aged.

In order to test and demonstrate aging effects in the conductance relaxation of our granular Al films, we have used the following protocol. The samples are quenched from room temperature to 4K, maintained there under a fixed gate voltage $V_{g0}$ until the gate voltage is changed to a new value $V_{g1}$ at time $t_e$. In a first set of measurements, we have measured at time $t_e + t_w$ the amplitude of the new conductance dip dug during $t_w$ at $V_{g1}$. As shown in Fig. 2, we clearly found a small but significant decrease of the dip amplitude as a function of $t_e$ ($t_e$ is the age of the system when $V_g$ is changed to $V_{g1}$). The older the system is, the longer it takes to dig a new dip, or in other words, the stiffness of the system increases with its age. This trend is confirmed by an other set of experiments where we follow the time conductance relaxation laws after the gate voltage is changed to $V_{g1}$. Deviations from a logarithmic time dependence are visible and the relaxation rate is found to depend on the age of the system, the younger the system, the faster the relaxation.

Surprisingly, most of the previous field effects experiments done on disordered insulators did not consider strictly speaking aging effects. "Simple aging" has often been reported [1,3] but it refers to the fact that, when submitted to an external perturbation (a $V_g$ or an electric field change) during $t_w$, the subsequent time relaxation curves of the conductance collapse on a single master curve when plotted as a function of $t/t_w$. The important point is that prior to the external perturbation, the sample is allowed to age for a long time after the quench at 4K ($t_e$ is of one day or more). "Simple aging" measurements were thus done in a regime such that $(t, t_w) << t_e$ and could not be sensitive to any "true aging" effect. Furthermore and contrary to our "true aging" results, the "simple aging" and its master curve shape could be explained by the trivial response of independent degrees of freedom relaxing back and forth under $V_g$ changes and having a flat distribution of the log of relaxation times [3].

Our findings strengthen the similarity between our system and other glasses. Qualitatively similar features are for example seen in the magnetic properties of spin glasses [12]. They also demonstrate the existence of non trivial dynamics and correlations in the electronic properties of our granular Al films. It would be now interesting to know if this aging also exists in other disordered insulators, such as indium oxide for example. We should mention that aging phenomena were found in numerical simulations on the Coulomb glass problem [13].

## 5. Conclusion

In granular Al thin films of micron size, we have observed reproducible mesoscopic conductance fluctuations as a function of gate voltage. We show that these fluctuations are not affected by the conductance dip that forms under a fixed gate voltage and which is also found in macroscopic samples. This result may be more readily interpreted in terms of an electron glass origin for the conductance dip, but further theoretical analysis would be needed to definitely rule out other scenarios. In macroscopic samples, we also demonstrate aging effects on the conductance relaxation following gate voltage changes. Like in other glassy systems, the relaxation is slower when the system has aged since cooling. This observation opens new issues such as memory and rejuvenation experiments as a function of temperature.

**Acknowledgements**

This work was supported by the French National Research Agency (contract N°ANR-05-JC05-44044).

Fig. 1: Normalized conductance for a microscopic granular Al sample (T = 4.2K, $R_{\square}$ = 27MΩ, channel size 20μm x 25μm). The gate voltage was changed from -15V to +15V after the spectra labeled t = 0s. Scans 400s long were taken every 2000s. The digging of a conductance dip is clearly visible around $V_g$ = +15V, while the background fluctuations remain essentially unchanged. Some noise is superimposed on the fluctuations pattern. The lower curves have been shifted for clarity.

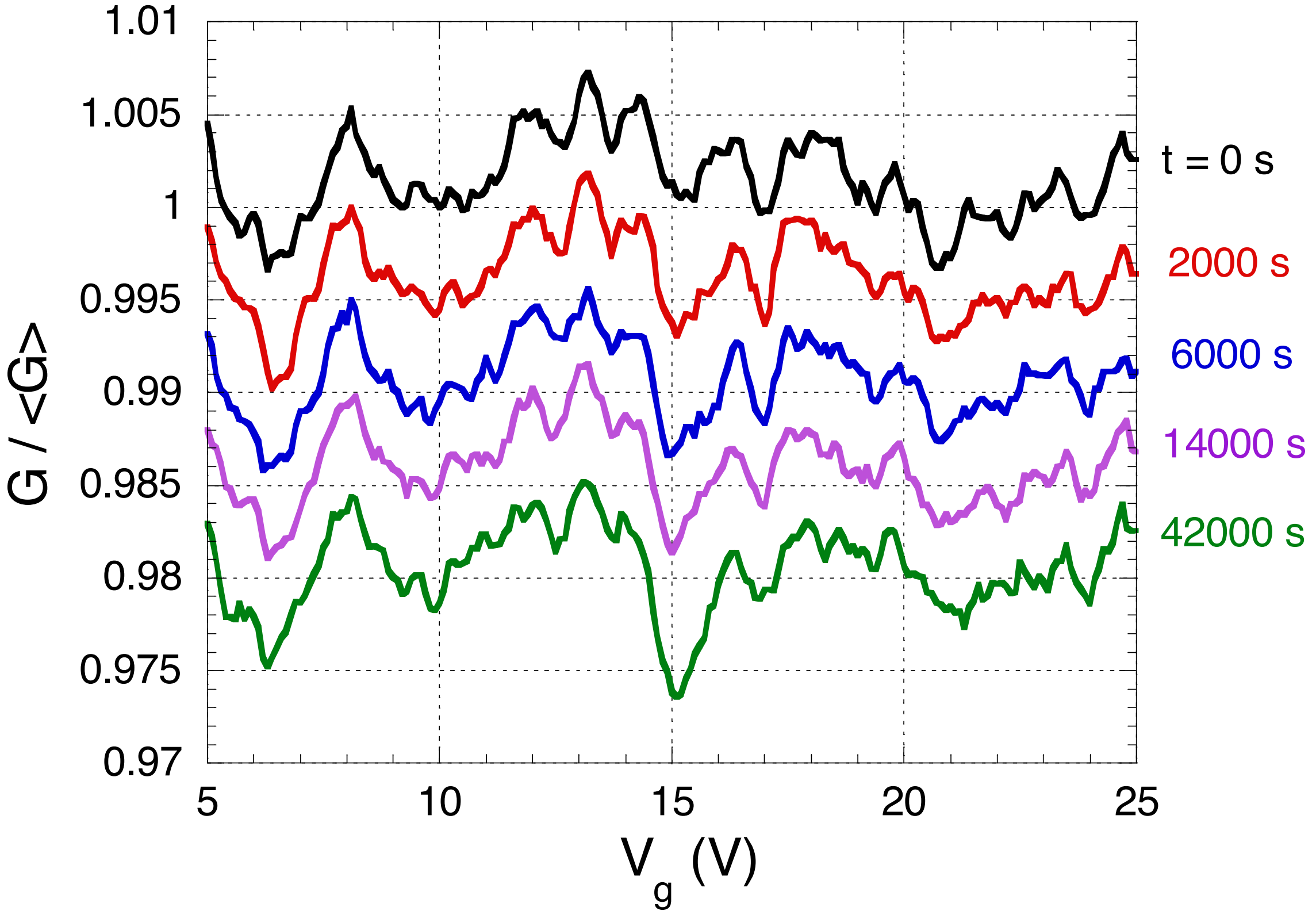

Fig. 2: Relative amplitude of a new dip dug during $t_w$ = 1350s as a function of the time $t_e$ elapsed since the quench at 4.2K ($R_\square$ = 8GΩ). The straight line is a guide for the eyes.

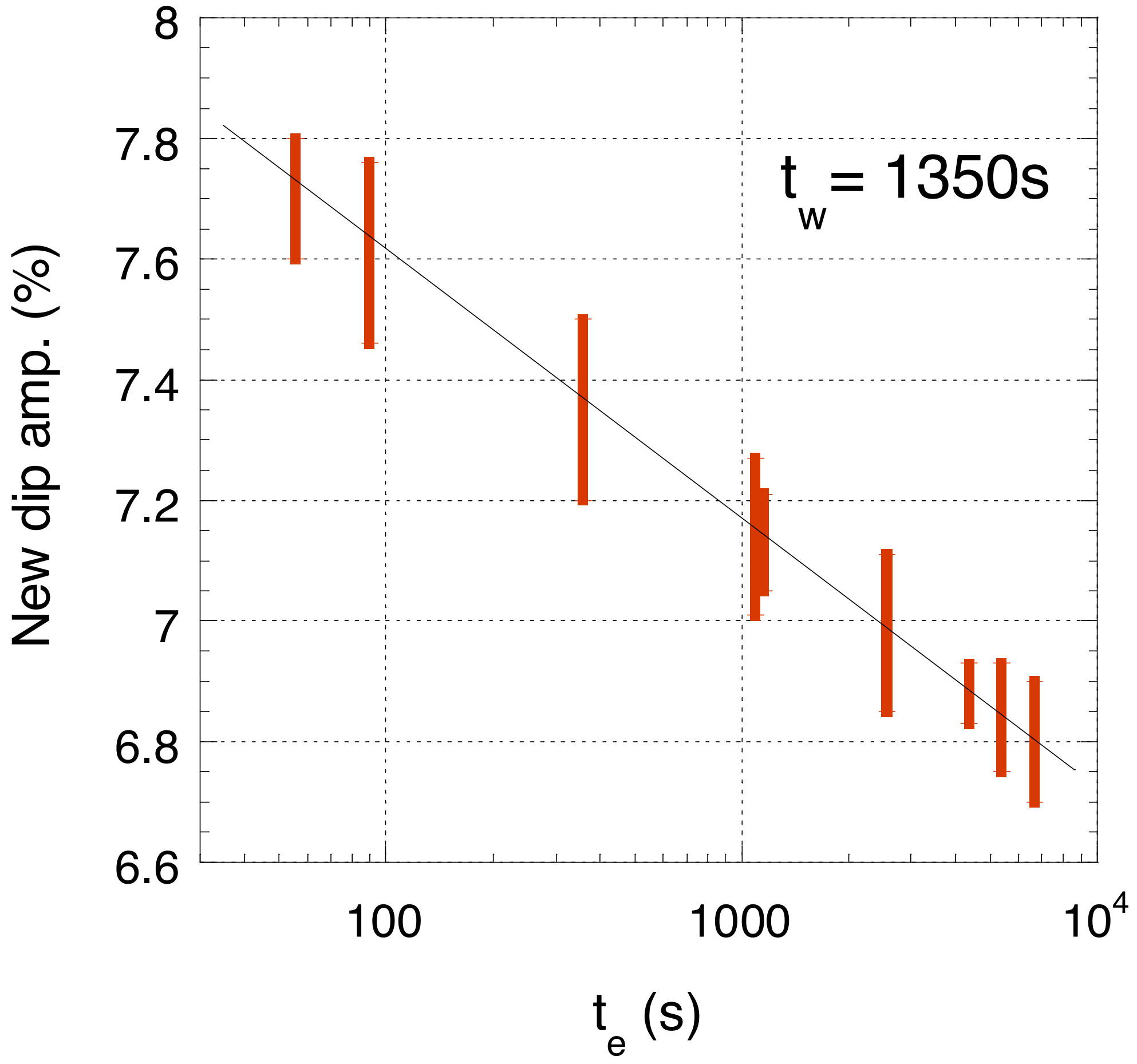